\documentclass[12pt,aps,preprint,tightenlines]{revtex4}

\usepackage{amsmath}
\usepackage{amsfonts}
\usepackage{epsfig}
\usepackage{amssymb}

\def\a{\alpha}

\def\g{\gamma}

\def\d{\delta}

\def\w{\omega}
\def\c{\nabla}

\begin{document}
\tighten

\title{Gravitational instability of Einstein-Gauss-Bonnet 
black holes under tensor mode perturbations}
\author{Gustavo Dotti and Reinaldo J. Gleiser}
\address{Facultad de Matem\'atica, Astronom\'{i}a y F\'{\i}sica,
Universidad Nacional de C\'ordoba, Ciudad Universitaria,
(5000) C\'ordoba, Argentina}
\email{gdotti@fis.uncor.edu}

\begin{abstract}
We analyze the tensor mode perturbations of static, spherically symmetric solutions 
of the  Einstein equations with a quadratic Gauss-Bonnet term in dimension $D > 4$. 
We show that the evolution equations for this type of perturbations can be cast in
a Regge-Wheeler-Zerilli form, and obtain the exact potential for the corresponding
Schr\"odinger-like stability equation. As an immediate application we prove that 
for $D \neq 6$ and $\alpha >0$, the sign choice for the Gauss-Bonnet 
coefficient suggested by string theory, 
all positive mass black holes 
of this type are stable. In the exceptional case $D =6$, we find a range of parameters 
where positive mass asymptotically flat black holes, with regular horizon, are unstable.
This feature is found also in general for $\alpha < 0$.
\end{abstract}

\maketitle

\section{Introduction}

Alternative gravity theories in higher dimensions have been attracting 
considerable attention, particularly the Einstein-Gauss-Bonnet (EGB) theory,
 which emerges as the low energy limit of string theory.
The EGB lagrangian is a linear combination of Euler densities continued 
from lower dimensions. It gives equations involving  up to 
second order derivatives of the metric,  and has the  same degrees 
of freedom as ordinary Einstein theory. A particular choice of  the coefficients 
in front of the  Euler densities gives theories where
the local Lorentz symmetry is  enlarged to a local $(A)dS$
symmetry \cite{ads1,ads2}. Interesting solutions to the EGB equations, many of 
them relevant to the development of the $AdS-CFT$ correspondence \cite{jm}, include 
a variety of black holes in asymptotically Euclidean or $(A)dS$ spacetimes. These 
solutions could be found mostly because they are highly symmetric.  
Analyzing their linear stability, however, confronts us with the high complexity of 
the EGB  equations, since the perturbative terms break the simplifying symmetries of 
the background metric. In this Letter we report on 
the stability of spherically symmetric, static 
solutions of the quadratic EGB theory.
These are preliminary results of ongoing  work on the stability 
of EGB black holes with arbitrary Einstein manifolds as horizons
\cite{bpe}

\section{Tensor perturbations of spherically symmetric EGB spacetimes}

The lowest order Einstein-Gauss-Bonnet (EGB) vacuum equations are
\begin{equation} \label{lovelock}
0 = {\cal{G}}_b{}^a \equiv \Lambda {G_{(0)}}_b{}^a + {G_{(1)}}_b{}^a 
+ \a {G_{(2)}}_b{}^a 
\end{equation}
Here $\Lambda$ is the cosmological constant, ${G_{(0)}}_{ab} = g_{ab}$ 
the spacetime metric, $ {G_{(1)}}_{ab} = R_{ab} -\frac{1}{2} R g_{ab}$
the Einstein tensor and  
\begin{multline}
{G_{(2)}}_b{}^a = R_{cb}{}^{de}  R_{de}{}^{ca}  -2 R_d{}^c R_{cb}{}^{da}
-2 R_b{}^c R_c{}^a + R  R_b{}^a 
 -\frac{1}{4} \d^a_b \left(
R_{cd}{}^{ef}R_{ef}{}^{cd} - 4 R_c{}^d R_d{}^c + R^2 \right)  
\end{multline}
the quadratic  Gauss-Bonnet tensor. These  are 
 the first  in a tower 
${G_{(s)}}_b{}^a, s=0,1,2,3,...$ of tensors  of order $s$ 
in $R_{ab}{}^{cd}$  constructed by Lovelock \cite{Lovelock}.
 As shown in  \cite{Lovelock},
the most general rank two, divergence 
free symmetric tensor that can be constructed out 
of the metric and its first two derivatives in a spacetime of dimension $d$, 
is a linear combination  of the ${G_{(s)}}_b{}^a$ with $2s \leq d$
\cite{Lovelock}.\\
 Here we  consider the spherically symmetric case, 
 a spacetime of dimension $D=n+2$ with  metric
\begin{equation} \label{bh}
ds^2 = -f(r) dt^2 + \frac{1}{f(r)} dr^2 + r^2 \bar g _{ij} dx^i dx^j,
\end{equation}
where $\bar g _{ij} dx^i dx^j$ is the line element of the
 unit $n-$dimensional sphere $S^n$.
We use indices $i,j,k,l,m,...$ and a  bar for tensors and operators on $S^n$, whereas  
$a,b,c,d,...$ are generic indices.
The nonzero components of the Riemann tensor of the metric 
(\ref{bh}) are
\begin{equation} \label{rie}
R_{tr}{}^{tr} = -\frac{f''}{2} , \hspace{.5cm} R_{it}{}^{jt} = R_{ir}{}^{jr}
= -\frac{f'}{2r} \d_i^j , \hspace{.5cm} R_{ij}{}^{kl} = \left( \frac{1-f}{r^2} \right)
(\d^k_i \d^l_j - \d^k_j \d^l_i)
\end{equation}
Inserting (\ref{rie}) in (\ref{lovelock}) we find that (\ref{bh}) solves the EGB equation if 
\begin{equation}
\label{fofr}
f(r) = 1 - r^2 \psi(r), 
\end{equation}
and $\psi(r)$ satisfies \cite{www} 
\begin{equation}
\label{psiofr}
\frac{\alpha n(n-1)(n-2)}{4} 
\psi(r)^2 + \frac{n}{2} \psi(r)  -\frac{\Lambda}{n+1}= \frac{\mu}{r^{n+1}}
\end{equation}
We consider {\em tensor} perturbations around (\ref{bh})
\begin{equation}
g_{ab} \to g_{ab} + h_{ab}.
\end{equation}
which 
 are those satisfying $h_{a b}=0$ unless 
$(a,b)=(i,j)$. Tensor perturbations are believed to be the only potentially unstable modes 
in ordinary Einstein theory \cite{gh}. We choose the gauge where $h_{ab}$ 
is transverse traceless. This is easily seen to imply that the restriction  
of $h_{ab}$  to the sphere 
is transverse traceless, and so can be expanded 
 using a basis of eigentensors
of the laplacian \cite{higu}. Thus,  we need only consider the case 
\begin{equation} \label{pert}
h_{ij}(t,r,x) = r^2 \phi(r,t) \bar h_{ij}(x) 
\end{equation}
where
\begin{equation} \label{pert1}
\bar \c^k \bar \c_k  \bar h_{ij}
= \g \bar h_{ij} ,\hspace{1cm} \bar \c^i  \bar h_{ij} = 0 , 
\hspace{1cm} \bar g ^{ij}  \bar h_{ij} = 0
\end{equation}
Solutions to equations (\ref{pert1}) are worked out in \cite{higu}, 
where it is shown that the spectrum of eigenvalues is $\gamma=-l(l+n-1)+2$,
$l=2,3,4,...$
The  components of the first  order 
variations $\d {G_{(s)}}_b{}^a,\; s=0,1,2$ under (\ref{pert}) are 
trivial unless $(a,b)=(i,j)$. After a long calculation the  $(i,j)$
 components are found to be
\begin{eqnarray}
\d {G_{(0)}}_i{}^j &=& 0 \\
\d {G_{(1)}}_i{}^j &=& \d R_i{}^j =\left[ \left( \ddot \phi - f^2 \phi'' \right)\frac{1}{2f} - \phi' \left(\frac{f'}{2} + \frac{nf}{2r} \right) + 
\frac{\phi}{2r^2} \left( 
2-\g \right) \right] \bar h_i{}^j 
\end{eqnarray}
and
\begin{multline}
\d {G_{(2)}}_i{}^j = \left\{ \left( \ddot \phi - f^2 \phi'' \right)  \left(\frac{n-2}{2 r^2 f} \right)  \left[ -r f' + (n-3)(1-f)  \right]  \right. \\ 
 \left. + \phi' \left(\frac{n-2}{2 r^3} \right) \left[ (n-3) \left((n-2)(f^2-f)-r f' \right) + r^2 (f'^2 + f'' f) +
(3n-7) r  f' f 
 \right]  \right. \\ 
\left.  + \phi \left(\frac{\g-2}{2 r^4} \right) \left[ r^2 f'' + 2 (n-3) r f' + (n-3)(n-4)(f-1) \right] 
 \right\} \bar h_i{}^j 
\end{multline}
Setting $\phi(r,t) = e^{\w t} \chi(r)$ the linearized EGB  equations
\begin{equation}
\d {G_{(1)}}_a{}^b + \a \d {G _{(2)}}_a{}^b = 0 
\end{equation}
around the metric (\ref{bh}) reduce to  a second order ODE 
for $\chi(r)$. By further introducing, 
\begin{equation}
\label{factor} 
\Phi(r) = \chi(r) K(r) 
\end{equation}
with,
\begin{equation}
\label{k} 
K(r) = r^{n/2-1} \sqrt{r^2+\alpha (n-2)\left((n-3)(1-f)-r \frac{df}{dr} \right)} 
\end{equation}
and switching to ``tortoise" coordinate $r*$, defined by ${dr*}/dr=1/f$, 
this ODE can be cast in the Schr\"odinger form,
\begin{equation} \label{s}
- \frac{d^2 \Phi}{dr*^2}  + V(r(r*)) \Phi = - \w ^2 \Phi \equiv E \Phi
\end{equation}
The solutions will therefore be stable if (\ref{s}) has no negative eigenvalues.
On the other hand, a negative eigenvalue ($E < 0)$ signals the possibility of an 
instability that requires also consideration of the normalization of the 
corresponding eigenfunctions (see, e.g. \cite{gh} for details). 

The explicit form of the potential $V(r)$ as a function of $r$ and the parameters of the 
theory is rather lengthy. We notice however that if we introduce the function,
\begin{equation} \label{v2}
q = \left(\frac{f (2-\gamma)}{r^2} \right) \left(  \frac{(1-\alpha f'')r^2 +\alpha (n-3) \;
 [(n-4) (1-f)-2r  f' ]}{r^2 +\alpha (n-2) \left[ (n-3)(1-f)-r f' \right]}
\right)
\end{equation}
the potential is given by,
\begin{equation}\label{v1}
V(r) = q(r) +\left(f \frac{d}{dr} \ln(K) \right)^2
 + f \frac{d}{dr} \left(f \frac{d}{dr} \ln(K) \right)
\end{equation}

Eqns (\ref{factor})-(\ref{v1}) are the main result of this paper, 
(\ref{v1}) being the exact potential of the Schr\"odinger-like 
stability equation for spherically symmetric EGB blackholes of arbitrary mass and 
cosmological constant.  Clearly, it can be applied to the cosmological 
solutions of the EGB equations that result by setting $\mu=0$. Moreover, our results are 
readily seen to reproduce those in \cite{gh} in the $\alpha=0$ (Einstein gravity) limit, which was also 
extensively studied by Kodama and Kodama and Ishibashi (see, e.g., \cite{ki} 
and references therein),
as well as 
the restricted cases studied in \cite{n} and \cite{bd}. In what follows, as an 
application of the formalism, we analyze briefly the case $\Lambda=0$ 
for general $n$, and also the $n=3$ and $n=4$ BTZ black holes \cite{ads2}.
The general case will be considered in a more extended version of this paper, 
currently in preparation \cite{bpe}.

\section{Stability of Einstein-Gauss-Bonnet black holes}

We recall that for $\Lambda=0$, on account of (\ref{fofr}) and (\ref{psiofr}), 
for asymptotically flat Einstein-Gauss-Bonnet black holes with regular horizon $f(r)$ 
takes the form \cite{www},
\begin{equation}
\label{fofr2}
f(r)= 1 +\frac{r^2}{(n-1)(n-2)\alpha}\left[1- 
\sqrt{1+\frac{4 (n-1)(n-2)\alpha \mu}{n r^{n+1}}}
\right]
\end{equation}
where $\mu >0$ corresponds to positive mass. 
We consider first $\alpha> 0$  which is  the relevant case for string motivated theories. 
Then, for any $\mu > 0$, there is a regular horizon at
$r=r_H$, and $f(r)$ grows monotonically from zero to one as 
$r$ grows from $r_H$ to infinity.
From (\ref{fofr2}), 
\begin{equation}
\label{muofrH}
\mu= n [ \alpha (n-1)(n-2)+ 2 r_H^2 ] r_H^{(n-3)}.
\end{equation}

Going back to (\ref{s}), a sufficient 
criterion for stability is that $V(r)$ is positive for $r> r_H$. If we consider (\ref{v1}),
we notice that the second term on the R.H.S. is positive definite in all cases, while, 
a long computation shows that the first and third terms are also positive definite for $r>r_H$, 
for $n =3$ and all $n>4$, so all these cases are stable under tensor perturbations. 
The $n=4$ case is exceptional. Here we notice that, since 
 $V(r(r*))$ is bounded in 
$-\infty < r* < +\infty$, 
with $V(r(r*)) \rightarrow 0$ for $r* \rightarrow \pm \infty$, a sufficient condition 
for the existence of a bound state of negative energy is \cite{coso},
\begin{equation}
\label{coso1}
\int_{-\infty}^{+\infty}V(r(r*)) dr* < 0
\end{equation}
This can be written as an integral over $r$,
\begin{equation}
\label{coso2}
\int_{r_H}^{+\infty}(V(r)/f(r)) dr < 0
\end{equation}
The second term on the right in (\ref{v1}), divided by $f$, is positive, 
while the
third, divided by $f(r)$, is a total derivative in $r$, and gives a vanishing contribution
on account of its behaviour for $r \rightarrow r_H$ and $r \rightarrow +\infty$, 
as is easily seen from (\ref{k}). The ``dangerous'' contribution comes then from
 $q(r)/f(r)$. In fact, since $q(r)$ contains the (positive) factor $(2-\gamma)$, which 
can be arbitrarily large for spherical horizons, while the other terms in (\ref{v1}) are 
independent of $\gamma$,
the condition,
\begin{equation}
\label{coso3}
\int_{r_H}^{+\infty}q(r)/((2-\gamma)f(r)) dr < 0
\end{equation}
implies that (\ref{coso1}) will be  satisfied for a sufficiently large $\gamma$. 
Note that  $\alpha$ has dimension $r^2$ and that for $n=4$, $\mu$ has  dimension $r^3$. 
Introducing $z \equiv \mu  \alpha^{-3/2}$ 
in (\ref{muofrH}) we find
\begin{equation} \label{h4} 
r_H =  \frac{\sqrt{\alpha}}{2}  \left[ \frac{ \left( 2 z
 + 2 \sqrt{
16 + z^2} \right)^{2/3} -4}{\left( 2 z + 2 \sqrt{
16 + z^2} \right)^{1/3}} \right]
\end{equation}
so that 
$r_H \to 0^+$  as $z \to 0^+$ ($\mu \to 0^+$). Setting $n=4$ in (\ref{v2}) 
and defining $x \equiv r (\mu \alpha)^{-1/5}$ gives
\begin{equation}
\label{coso4}
\frac{q}{(2-\gamma) f}= (\mu \alpha)^{-2/5} \left[ \frac{2 (x^5 +6)^2 - 75}{
2 x^2 (x^5+6)(x^5+1)}\right]   
\end{equation} 
The integral in (\ref{coso3}) can be given in closed form using (\ref{coso4}),
 but the 
expression is too long and difficult to handle.
We may however show that the integral in (\ref{coso3}) is negative
if we first change variables to $u=1/x$
\begin{equation}
\int_{r_H}^{+\infty} \frac{q(r)}{(2-\gamma)f(r)} dr = \frac{1}{(\mu \alpha)^{1/5}} 
\int_0^{1/x_H} \left[ \frac{2(1+6u^5)^2-75u^{10}}{2(1+6u^5)(1+u^5)} 
\right] du,
\end{equation}
and then note  from (\ref{h4}) that  $x_H \simeq \mu^{4/5} \alpha^{-6/5}/12$ for $\mu \gtrsim 0$, which implies that 
the upper limit of the R.H.S.  integral above 
tends to infinity as $\mu \to 0^+$. Since the integrand stabilizes in 
$-1/4$ for large $u$, the integral is certainly negative
for sufficiently small 
$\mu$.
To illustrate this point we display in  Figure 1   the
potential $V(r(r*))$ as a function of $r*$ for two $\alpha=1$ cases.
Figure 1.a shows  the potential 
corresponding to a small $\mu$ case where (\ref{coso1}) holds, whereas Figure 1.b 
shows the  potential of a large $\mu$ solution. This is positive definite 
and therefore  does not allow bound states. \\

In closing this section we remark that, in space time dimensions $D=5$ and $D=6$, EGB 
black holes 
with a cosmological constant contain as particular cases the corresponding BTZ black holes 
\cite{ads2}. In the notation of this Letter and that of \cite{ads2} we have,
\begin{eqnarray}
\alpha & = & \ell^2/2\;,\; \Lambda = -3/\ell^2 \;,\; \mu =3 \ell^2 (M+1)/4 
\;\;,\;(\mbox{for}\; D =5 ) \nonumber \\ 
\alpha & = & \ell^2/6\;,\; \Lambda = -5/\ell^2 \;,\; 
\mu =2 \ell^2 M \;\;,\;(\mbox{for}\; D =6)   
\end{eqnarray}

Interestingly, we find that all $D=5$ solutions are stable, 
while {\em all} solutions 
are unstable for $D=6$. We recall that these cases were actually excluded in the analysis
in \cite{ads2}, on considerations based on cosmic censorship.

\section{Comments and conclusions}

Summarizing the results reported in this Letter, we have found an explicit form
for the Schr\"odinger-like equation governing the evolution of linear tensor 
perturbations of static spherically symmetric solutions of EGB vacuum equations. 
As a first application we proved the stability of (asymptotically flat) EGB black holes
with positive mass and coupling constant $\alpha$, in dimension $D = n +2$, for $n=3$, 
and $n > 4$. In the case $n=4$ we found the unexpected result that the EGB black holes are 
stable only for sufficiently large mass. The nature of the instability of the small mass
black holes is an intriguing question, outside the scope of the present work
(a thermodynamic instability of some asymptotically $(A)dS$ EGB black holes 
was also found in \cite{npb}). 
 Preliminary results
indicate that in the
 $\alpha < 0$ case, for all $n \geq 3$ there are solutions that represent static 
black holes 
with regular horizons, that are, however, unstable under tensor perturbations. The results 
obtained in this Letter are straightforwardly  extended to blackholes 
with non positive constant curvature horizons, as those 
studied in \cite{atz}. These are currently being 
analyzed  together
with other black holes having more general manifolds as horizons.

\section*{Acknowledgments}

We are grateful to Ricardo Troncoso, Jorge Zanelli and Jorge Pullin  for 
a number of useful comments on a preliminary version of this paper.
This work was supported in part by grants of the National University of
C\'ordoba and Agencia C\'ordoba Ciencia (Argentina). It was also supported in
part by grant NSF-INT-0204937 of the National Science Foundation of the US. The
authors are supported by CONICET (Argentina).

\begin{figure}
\includegraphics[width=6in]{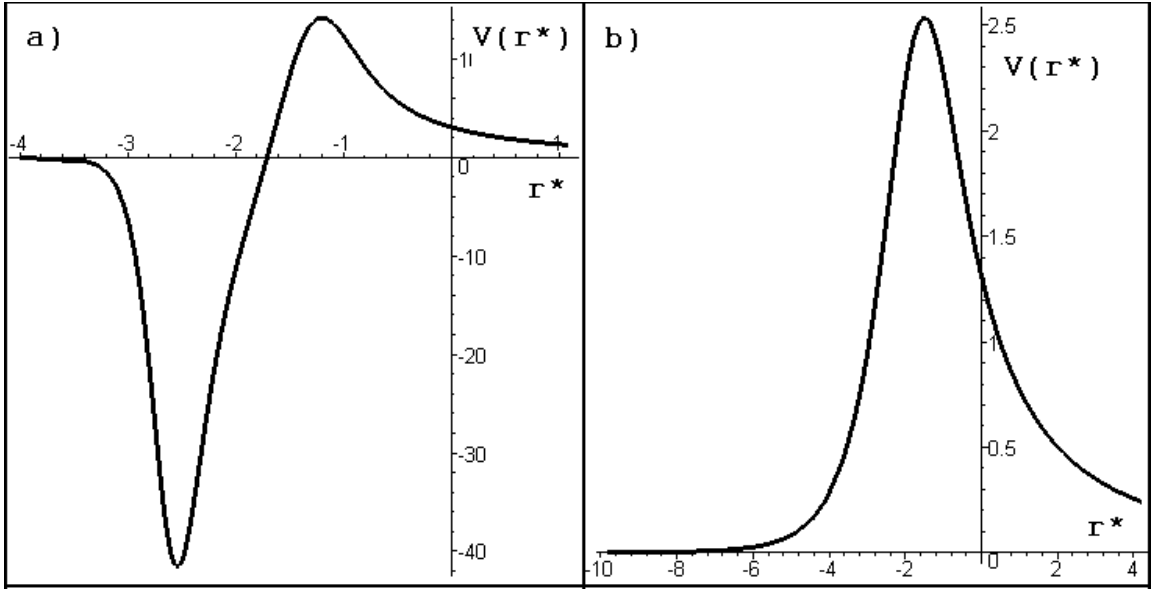}
\caption{\label{fig1} The potential $V(r(r^*))$ as a function of $r^*$. 
Figure 1 a) 
corresponds to $\mu = 0.4$, $\alpha = 1$, while for Figure 1 b) we have 
taken 
$\mu = 8$, $\alpha = 1$. The values of the integral in (\ref{coso1}) are 
$-9.008...$ 
for Figure 1 a), and $+10.39...$ for Figure 1 b). }
\end{figure}

\end{document}